\documentclass[12pt,a4paper]{article}
\usepackage{epsfig}
\begin{document}
\title{Single production of heavy top quark from the \\three-site Higgsless model}
\author{Chong-Xing Yue, Li-Hong Wang, Jia Wen\\
{\small  Department of Physics, Liaoning Normal University, Dalian
116029, China}\thanks{E-mail:cxyue@lnnu.edu.cn}\\}
\date{\today}

\maketitle
\begin{abstract}
We consider single production of the heavy top quark T predicted by
the three-site Higgsless model in future high energy collider
experiments, such as the high energy linear $e^{+}e^{-}$ collider
$(ILC)$, the linear-ring type $ep$ collider $(THERA)$, and the
$CERN$ Large Hadron Collider $(LHC)$. Our numerical results show
that the possible signals of the heavy top quark T might be observed
via the process $e^{+}e^{-}\rightarrow t\overline{T}\rightarrow
t\bar{t}Z$ in future $ILC$ experiments and via the subprocess
$qb\rightarrow q'T$ at the $LHC$.

\end{abstract}

\newpage
\noindent{\bf I. Introduction}

 To completely avoid the problems arising from the elementary Higgs scalar
  field in the standard model $(SM)$, various kinds of models for electroweak
 symmetry breaking $(EWSB)$ have been proposed, among which Higgsless
 models [1,2] are attractive because they provide a possible $EWSB$
 mechanism without introducing an elementary scalar Higgs
 boson. In this kind of models, $EWSB$ can be achieved via employing
 gauge symmetry breaking by boundary conditions in a higher
 dimensional theory space [3], and the unitarity of
 longitudinally-polarized $W$ boson and $Z$ boson scattering is preserved
 by the exchange of new vector gauge bosons [4]. Higgsless models open up
 new possibilities for new physics beyond the $SM$, which might
 produce observed signatures in future high energy collider
 experiments.

 Higgsless theories have been proposed in five warped dimensions
 [1,5], in five flat dimensions with brace kinetic terms [2], in six
 dimensions [6], and in four-dimensional theory space [7]. Since all
 of these theories are only low energy effective theories, their
 properties may be conveniently studied using deconstruction [8],
 which is a technique to build a four-dimensional gauge theory with
 an appropriate gauge symmetry breaking pattern. Deconstructed
 Higgsless models [7,9,10] have been used as tools to compute the
 general properties of Higgsless theories and to illustrate the
 phenomenological properties of this kind of new physics model
 beyond the $SM$.

 The simplest deconstructed Higgsless model incorporates only three
 sites on the deconstructed lattice, which is called the three-site
 Higgsless model [11]. In this model, the ordinary fermions are
 ideally delocalized, which preserves the characteristic of
 vanishing precision electroweak corrections up to subleading order
 [12]. Furthermore, the three-site model is sufficiently rich to
 describe the physics associated with fermion mass generation
 including the ordinary fermions and their heavy partners.

 Although differing in details, all of the Higgsless models have
 general features and collider phenomenology [13]. Because Higgsless
 models with ideally delocalized fermions have vanishing precision
 electroweak observables, it is necessary to look elsewhere for
 experimental signatures of this type of Higgsless models. Thus, in this
 paper, we will consider production and decay of the heavy top quark $T$
 coming from the three-site Higgsless model in future high energy
 collider experiments. In order to present our calculation, we state
 briefly the ingredients of this model.

 The three-site Higgsless model [11] has a standard color group and
 an extended $SU(2)\times SU(2)\times U(1)$ electroweak gauge
 group. It incorporates a
 $SU(2)\times SU(2)\times U(1)$ gauge group
 with coupling constants $g_{0}$, $g_{1}$, and $g_{2}$ respectively, 2
 nonlinear $(SU(2)\times SU(2))/SU(2)$ sigma models, in which
 the global symmetry groups in adjacent sigma models are identified
 with the corresponding factors of the gauge group. In the limit
 $g_{0}\ll g_{1}$ and $g_{2}\ll g_{1}$, the three-site
 Higgsless model approximately denotes the $SM$, and therefore there
 are:
\begin{equation}
g_{0}\simeq g =\frac{e}{S_{W}},\hspace{2.5cm}g_{2}\simeq
g'=\frac{e}{C_{W}},
\end{equation}
where $S_{W}=\sin\theta_{W}$ and $\theta_{W}$ is the Weinberg angle.

A fermion field in a general compactified five-dimensional theory
gives rise to a tower of $KK$ modes, the lightest of which can be
identified with the ordinary fermion. Thus, the fermions in the
three-site Higgsless model include the ordinary fermions and their
 heavy partners. The couplings of the heavy partner $T$ of the top quark to ordinary
particles, which are related to our calculation, can be written as
[11]:
\begin{eqnarray}
g_{L}^{ZtT} &=&
\frac{e}{2\sqrt{2}S_{W}C_{W}(\varepsilon_{tR}^{2}+1)}\chi
P_{L},\hspace{1.0cm} g_{R}^{ZtT} = \frac{e
\varepsilon_{tR}}{2S_{W}C_{W}(\varepsilon_{tR}^{2}+1)}P_{R};
\\g_{L}^{WTb} &=&
\frac{e(1-\varepsilon_{tR}^{2})}{2\sqrt{2}S_{W}(\varepsilon_{tR}^{2}+1)}\chi
P_{L},\hspace{1.6cm}g_{R}^{WTb} = 0.
\end{eqnarray}
Where $\varepsilon_{tR}=\lambda_{t}'/\tilde{\lambda}$ is a
flavor-dependent free parameter, in which $\lambda_{t}'$ and
$\tilde{\lambda}$ are the corresponding Yukawa coupling constants.
$\chi=g_{0}/g_{1}$ is a small free parameter. $P_{L}$ and $P_{R}$
are the left- and right- handed projection operators, respectively.

In the following section, we will use above equations to calculate
the cross sections for single production of the heavy top quark $T$
in the high energy linear $e^{+}e^{-}$ collider $(ILC)$, the
linear-ring type $ep$ collider $(THERA)$, and the $CERN$ Large
Hadron Collider $(LHC)$. Some phenomenological analysis are also
given.

\noindent{\bf II. Single production of the heavy top quark $T$ in
future high energy collider \hspace*{0.7cm}experiments }

Before giving numerical results, we need to specify the relevant
$SM$ parameters. These parameters are $m_{t}=171.4GeV$ [14],
$\alpha_{e}=1/128.8$, $\alpha_{s}=0.118$, $S^{2}_{W}=0.2315$,
$m_{W}=80.425GeV$ [15]. Except for these $SM$ input parameters, the
production cross section of the heavy top quark $T$ predicted by the
three-site Higgsless model is dependent on the free parameters
$\varepsilon_{tR}$, $\chi$, and $M_{T}$, in which $M_{T}$ is the
mass of the heavy top quark $T$ and its value has been discussed in
$Ref.$[16]. Considering the constraints of the electroweak
parameters $S$ and $T$ on the three-site Higgsless model [11,12], we
will assume $0.1\leq \varepsilon_{tR}\leq 0.5$, $0.06 \leq\chi\leq
0.3$, and $1TeV\leq M_{T}\leq2.5TeV$ in our numerical estimation.

\begin{figure}[htb]
\begin{center}
\epsfig{file=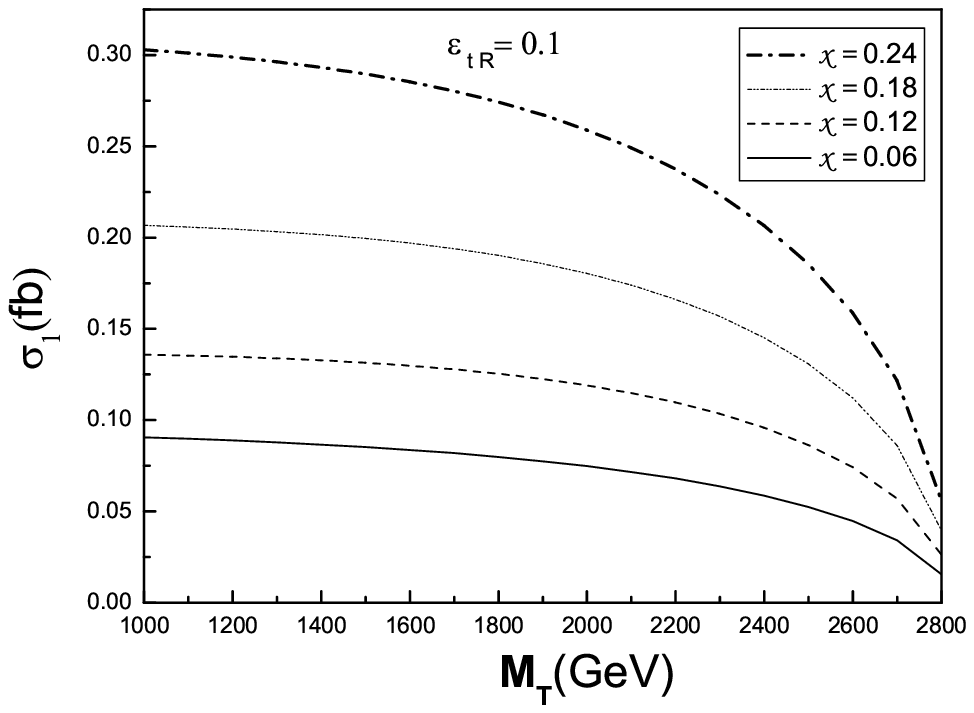,width=200pt,height=180pt}
\put(-110,5){(a)}\put(115,5){ (b)}\vspace{-0.25cm}
\epsfig{file=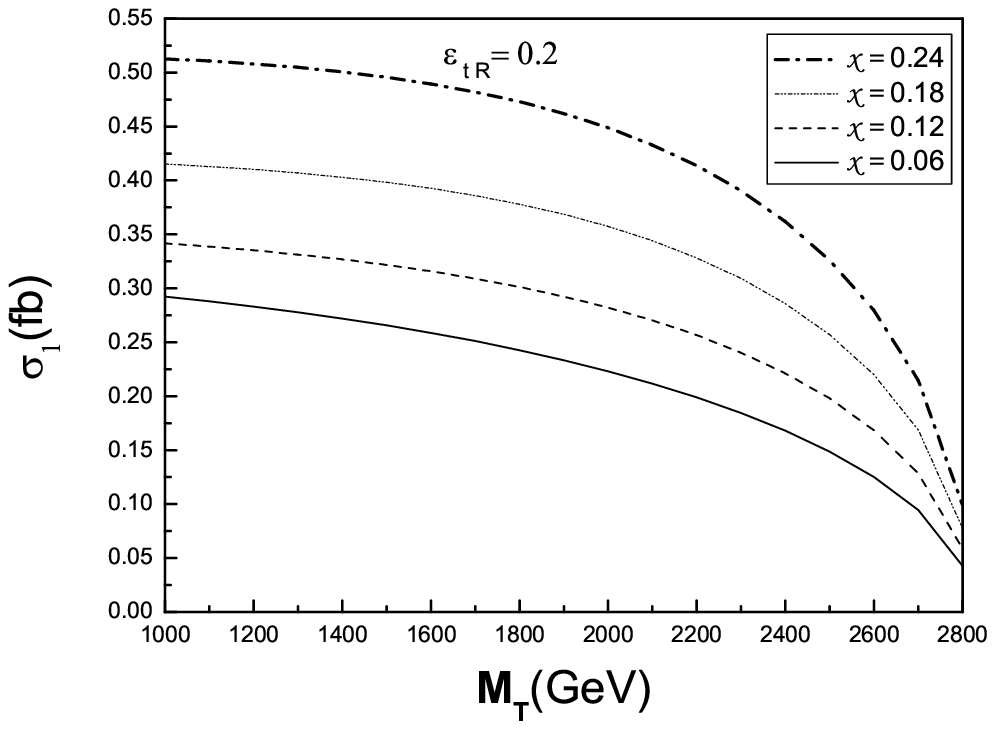,width=200pt,height=180pt} \hspace{-0.5cm}
\hspace{10cm}\vspace{-1cm}
\epsfig{file=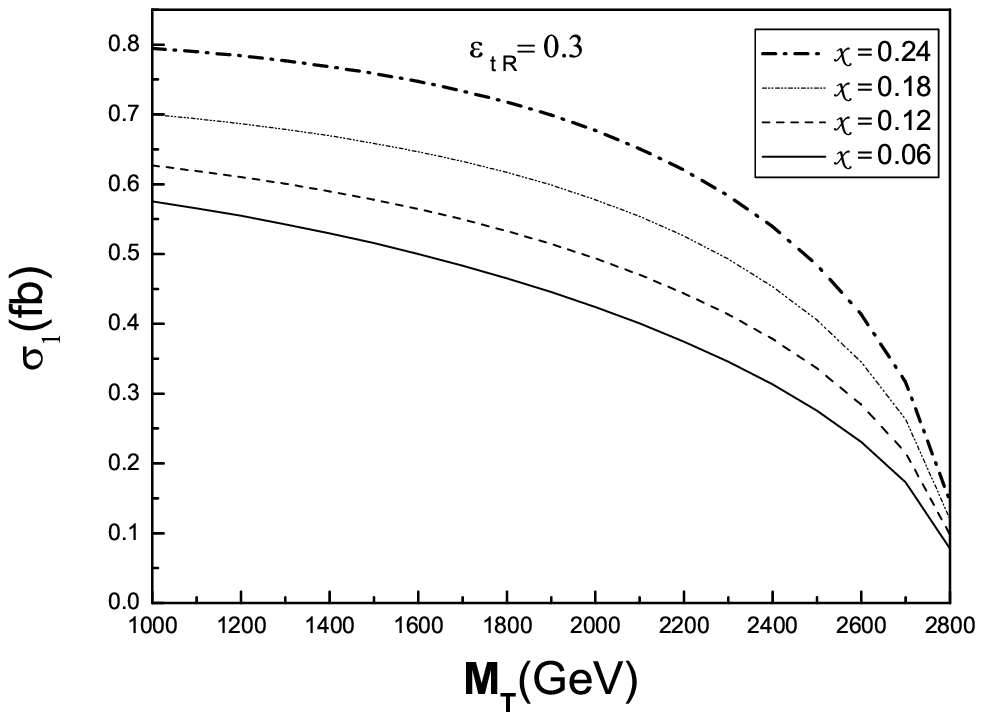,width=200pt,height=180pt} \hspace{-0.5cm}
\vspace{-0.25cm} \put(-110,5){(c)}\put(115,5){ (d)}
\epsfig{file=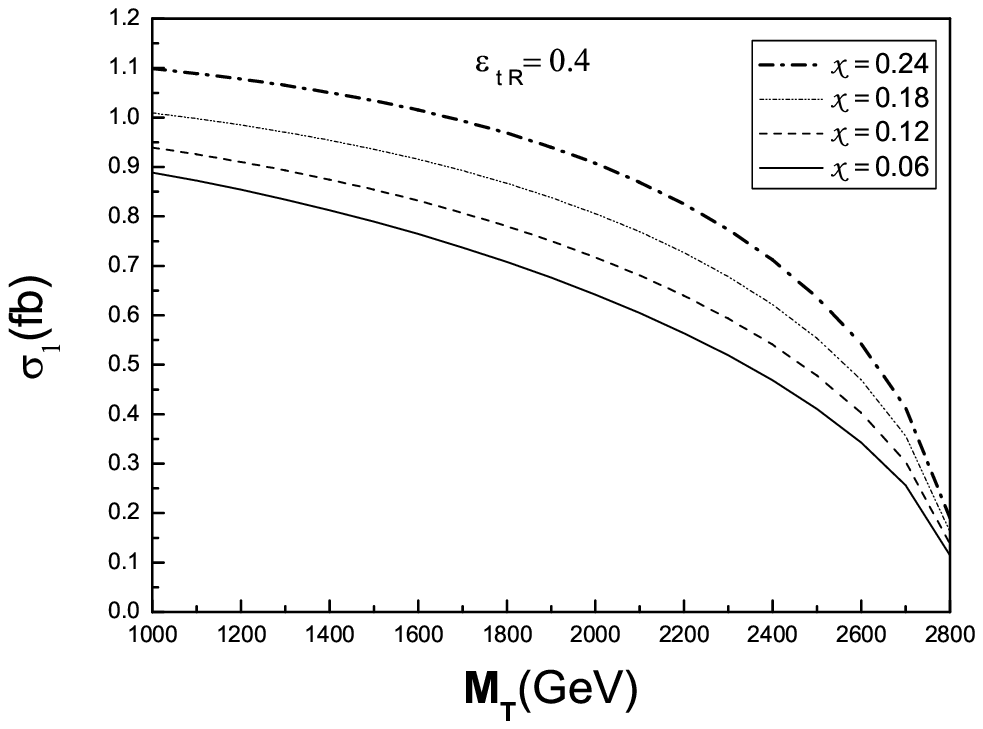,width=200pt,height=180pt} \hspace{-0.5cm}
\hspace{10cm}\vspace{-0.2cm} \caption { The cross section
$\sigma_{1}$ for the process $e^{+}e^{-}\rightarrow t\overline{T}$
as a function of the mass \hspace*{1.7cm} parameter $M_{T}$ for
$\sqrt{S}=3TeV$ and different values of the free parameters
\hspace*{1.7cm} $\chi$ and $\varepsilon_{tR}$.}
\end{center}
\end{figure}

From $Eq.(2)$ we can see that the heavy partner $T$ of the top quark
$t$ can be produced associated with a top quark $t$ via the
s-channel electroweak gauge $Z$ exchange in future $ILC$
experiments. Our numerical results are shown in $Fig.1$, in which we
have plotted the production cross section $\sigma_{1}$ for the
process $e^{+}e^{-}\rightarrow t\overline{T}$ as a function of the
$T$ mass $M_{T}$ for the center-of-mass $(CM)$ energy
$\sqrt{S}=3TeV$ and different values of the free parameters $\chi$
and $\varepsilon_{tR}$. One can see from $Fig.1$ that, in most of
the parameter space, the value of the cross section $\sigma_{1}$ is
 smaller than $1.2fb$. For example, for $0.06\leq\chi\leq0.24$, $1TeV\leq M_{T}
\leq2.8TeV$, and $\varepsilon_{tR}=0.4$, the cross section
$\sigma_{1}$ is in the range of $0.12fb\sim1.1fb$. Thus, there might
be several tens up to hundreds of $t\overline{T}$ events to be
produced at the $ILC$ with $\sqrt{S}=3TeV$ and a yearly integrated
luminosity of $\pounds=500fb^{-1}$[17].

To leading order, the heavy quark $T$ mainly decays into $tZ$ and
$Wb$ modes and their decay widths can be written as:
\begin{eqnarray}
\Gamma(T \rightarrow tZ) &\simeq& \frac{\sqrt{2}e^{2}\chi
\varepsilon_{tR}}{64\pi S^{2}_{W}
C^{2}_{W}(\varepsilon_{tR}^{2}+1)^{2}}M_{T} ,
\\\Gamma(T \rightarrow Wb) &\simeq& \frac{e^{2}(1-\varepsilon_{tR}^{2})^{2}
\chi^{2}}{128\pi
S^{2}_{W}(\varepsilon_{tR}^{2}+1)^{2}}M_{T}.
\end{eqnarray}

In terms of observability, the branching ratio $Br(T\rightarrow tZ)$
is larger than the branching ratio $Br(T\rightarrow Wb)$ and their
values are dependent on the values of the free parameters $\chi$,
$\varepsilon_{tR}$ and $M_{T}$. In most of the parameter space, the
value of $Br(T\rightarrow tZ)$ is about $85\%$. Thus the maximal
value of the production cross section for the process
$e^{+}e^{-}\rightarrow t\bar{t}Z$ contributed by the heavy top quark
$T$ can reach $1fb$. If we would like to detect possible signals of
the new heavy  $T$ quark via the decay channel $T\rightarrow tZ$,
the $SM$ background process $e^{+}e^{-}\rightarrow ttZ$ must be
considered, which has been extensively studied in literature (for
example, see [18]). It has been shown that appropriate cuts on the
$SM$ background process $e^{+}e^{-}\rightarrow t\bar{t}Z$ can
strongly reduce its cross section. Furthermore, the the $SM$
 process $e^{+}e^{-}\rightarrow t\bar{t}Z$ and the process
 $e^{+}e^{-}\rightarrow t\overline{T}\rightarrow
t\bar{t}Z$ are in completely different kinematics. Thus, the
possible signals of the heavy top quark $T$ might be detected via
the process $e^{+}e^{-}\rightarrow t\overline{T}\rightarrow
t\bar{t}Z$ in future $ILC$ experiments. If we assume that the heavy
top quark $T$ decays to $Wb$, then the $SM$ backgrounds mainly come
from the process $e^{+}e^{-}\rightarrow t\bar{t}$. Because large
production cross section for the process $e^{+}e^{-}\rightarrow
t\bar{t}$ and much small production cross section for the signal
process $e^{+}e^{-}\rightarrow t\overline{T}\rightarrow t\bar{b}W$,
the possible signals of the heavy top quark $T$ is more difficult to
be detected via the decay channel $T\rightarrow Wb$  than via the
decay channel $T\rightarrow tZ$.

\begin{figure}[htb]
\begin{center}
\epsfig{file=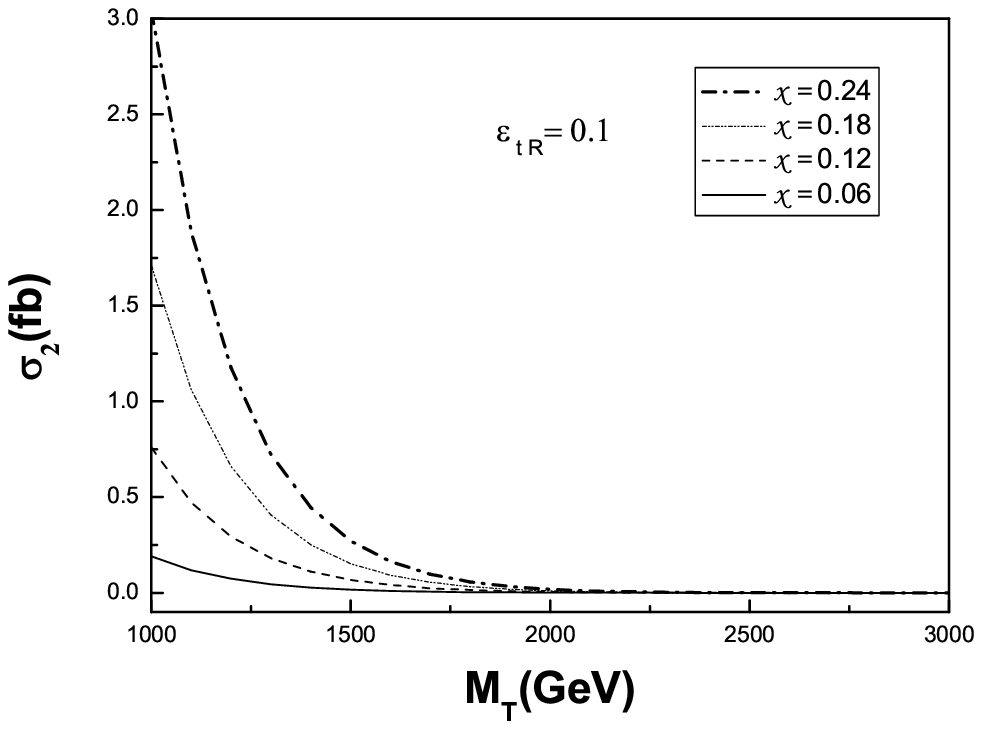,width=200pt,height=180pt} \put(-110,5){
(a)}\put(115,5){ (b)} \vspace{-0.25cm}
\epsfig{file=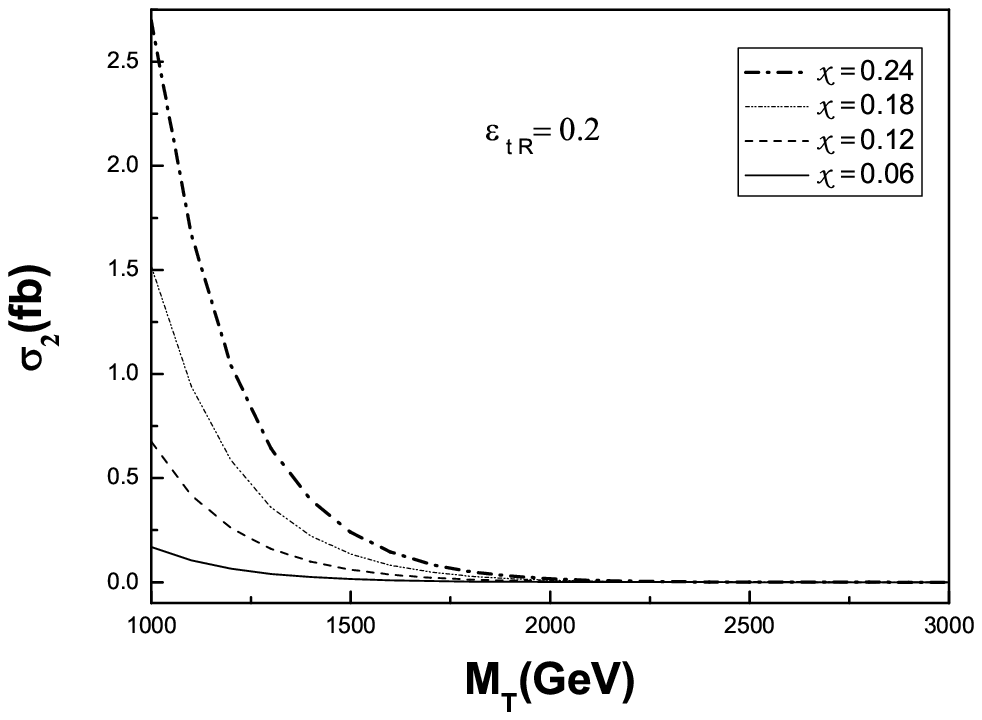,width=200pt,height=180pt} \hspace{-0.5cm}
\hspace{10cm}\vspace{-1cm}
\epsfig{file=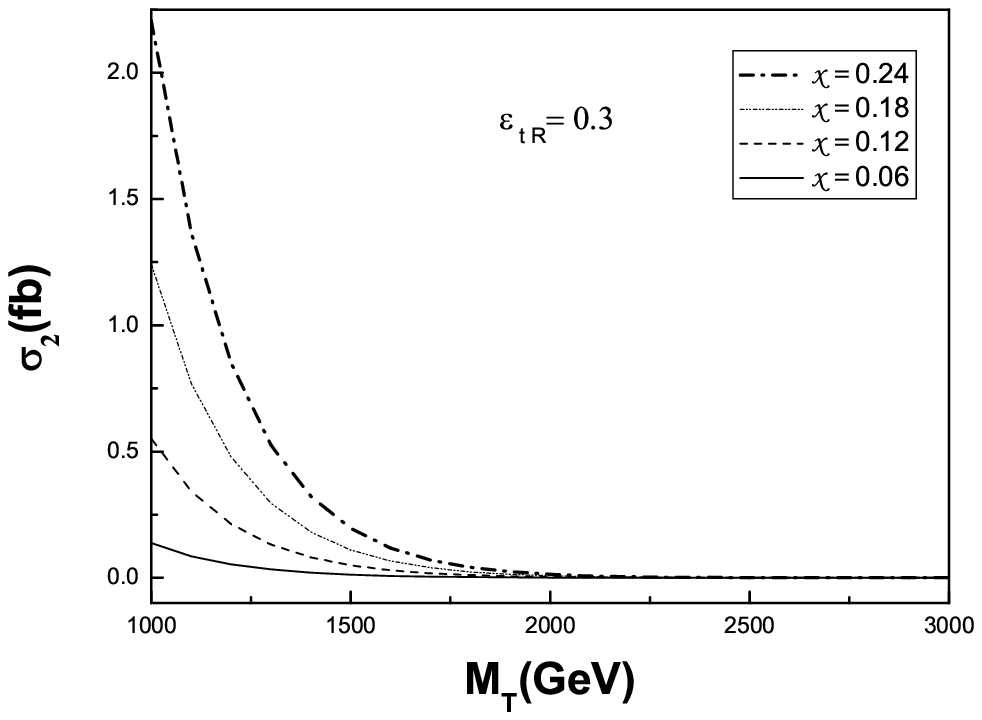,width=200pt,height=180pt} \hspace{-0.5cm}
\vspace{-0.25cm} \put(-110,5){(c)}\put(115,5){ (d)}
\epsfig{file=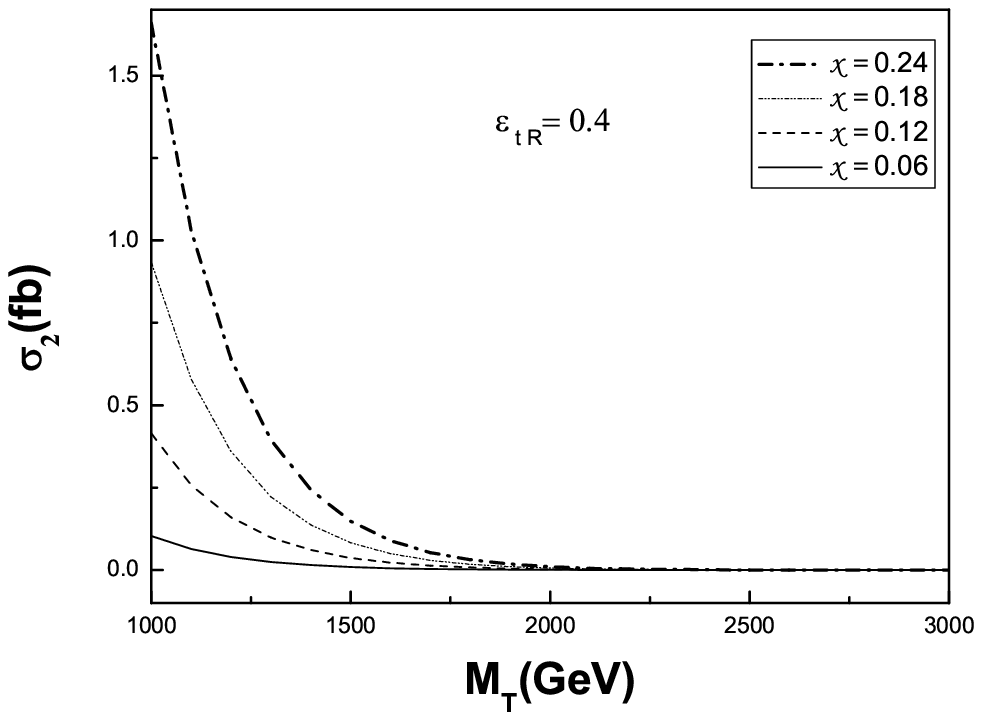,width=200pt,height=180pt} \hspace{-0.5cm}
\hspace{10cm}\vspace{-0.2cm} \caption { Same as Fig.1 but for the
process $ep \rightarrow ebX \rightarrow \nu_{e}TX$}
\end{center}
\end{figure}

The heavy top quark $T$ can be produced associated with an electron
neutrino via the t-channel $W$ exchange at the $THERA$ [19]. The
production cross section $\sigma_{2}$ for the process $ep\rightarrow
ebX\rightarrow\nu_{e}TX$ at the $THERA$ with the $CM$ energy
$\sqrt{S}=3.7TeV$ and the integral luminosity
$\pounds\simeq100Pb^{-1}$ [20] is shown in $Fig.2$, in which we plot
the cross section $\sigma_{2}$ as a function of the $T$ mass $M_{T}$
for different values of the free parameters $\chi$ and
$\varepsilon_{tR}$. In our numerical calculation, we have taken the
$CTEQ6L$ parton distribution function $(PDF)$ [21] for $b$ quark
$PDF$. From $Fig.2$, we can see that, in most of the parameter
space, the cross section $\sigma_{2}$ for single production of the
heavy top quark $T$ at the $THERA$ is larger than that at the $ILC$.
For $1TeV\leq M_{T}\leq3TeV$, $0.06\leq\chi\leq0.24$, and
$0.1\leq\varepsilon_{tR}\leq0.4$, the value of the cross section
$\sigma_{2}$ is in the range of $3.1fb\sim0.1fb$.

The decay modes $tZ$ and $Wb$ might provide characteristic
signatures for the discovery of the $T$ quark at the $THERA$.
Specially, the decay mode $Wb$ can generate nice signal event with a
b-jet, a charged lepton, and large missing energy, which can yield a
distinct experimental signature. However, the number of the signal
event b-jet+$l+\not{E}$ is too little to be detected in future
$THERA$ experiment with $\pounds\simeq100Pb^{-1}$. Certainly,
enhancing the value of the integral luminosity of the $THERA$
experiments can largely increase the number of the $ \nu_{e}T$
events. Thus, the possible signals of the heavy top quark $T$ might
be detected in  future $THERA$ experiment with high integral
luminosity.

At the leading order, the heavy top quark $T$ can be produced via
the partonic processes $q\overline{q'}\rightarrow T\bar{b}$,
$qb\rightarrow T q'$, and $gb\rightarrow TW$ at the $LHC$, as shown
in $Fig.3$. $q$ and $q'$ are the light quarks $u$, $c$, $d$, or $s$.

\vspace{0.5cm}
\begin{figure}[htb]
\begin{center}
\vspace*{-5cm}
\hspace*{-2.5cm}
\epsfig{file=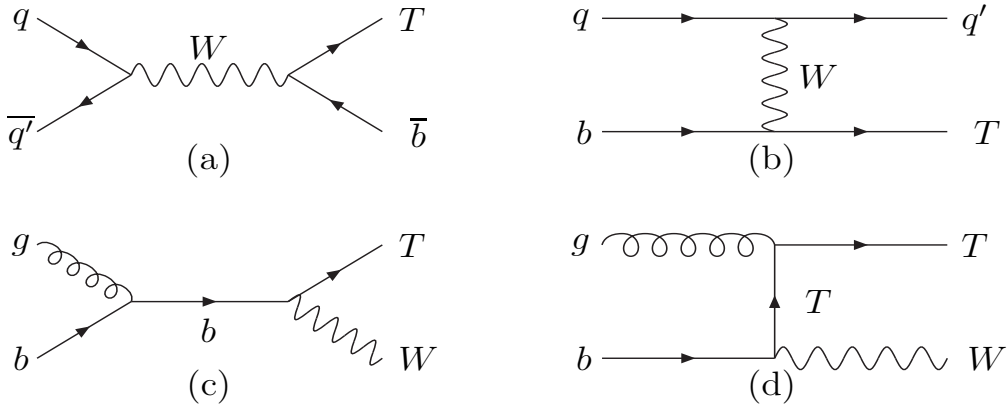,width=700pt,height=900pt} \vspace{-22cm}
\caption{Feynman diagrams for single production of the heavy top
quark $T$ at the $LHC$.}
\end{center}
\end{figure}
\vspace{0.2cm}

Since the subprocess $q\overline{q'}\rightarrow T\bar{b}$ proceeds
via the s-channel $W$ exchange with highly virtual propagator, the
production cross section of the $T$ quark associated with a $b$
quark is very small at the $LHC$. Our calculation results show that
its value is smaller than $5\times10^{-2}fb$ in most of the
parameter space. For the t-channel subprocess $qb\rightarrow q'T$,
it is not this case. Its production cross section $\sigma_{3}$ is
plotted in $Fig.4$ as a function of the $T$ quark mass $M_{T}$ for
different values of the free parameters $\varepsilon_{tR}$ and
$\chi$. One can see from $Fig.4$ that, for $\varepsilon_{tR}\leq
0.2$, $M_{T}\leq2TeV$, and $\chi>0.15$, the value of the production
cross section $\sigma_{3}$ is larger than $1fb$. If we assume
$\varepsilon_{tR}=0.1$, $\chi=0.18$, and $M_{T}=1.0TeV$, its value
can reach $22.5fb$. Thus, there will be several tens and up to
thousands of the $q'T$ events to be generated one year at the $LHC$
with $\sqrt{S}=14TeV$ and $\pounds=300fb^{-1}$.
\begin{figure}[htb]
\begin{center}
\epsfig{file=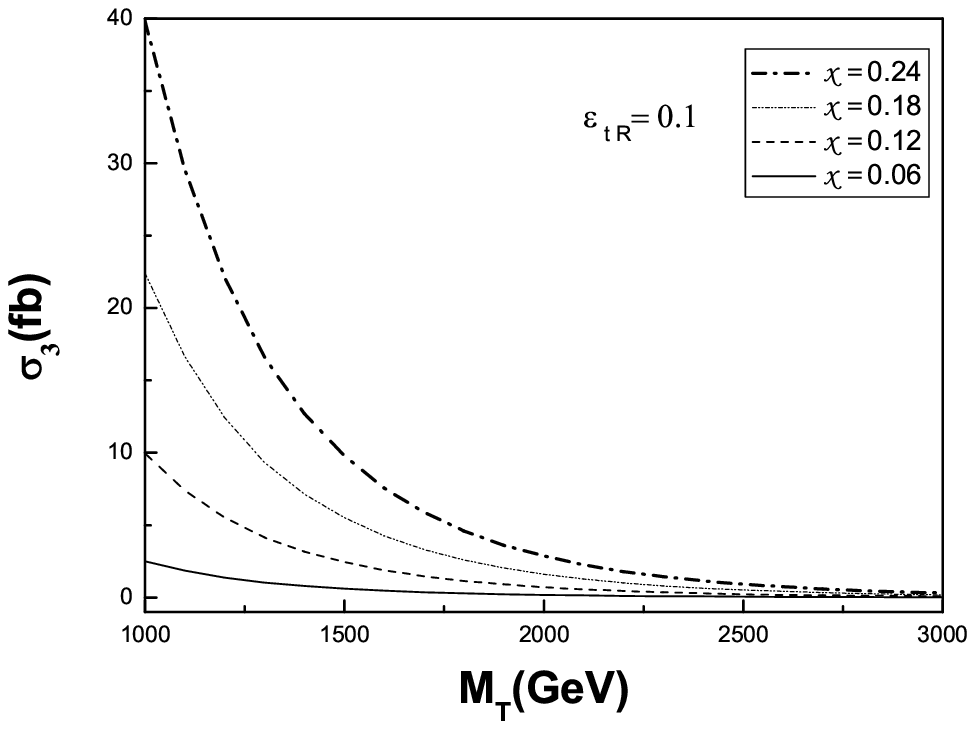,width=200pt,height=180pt} \put(-110,5){
(a)}\put(115,5){ (b)}\vspace{-0.25cm}
\epsfig{file=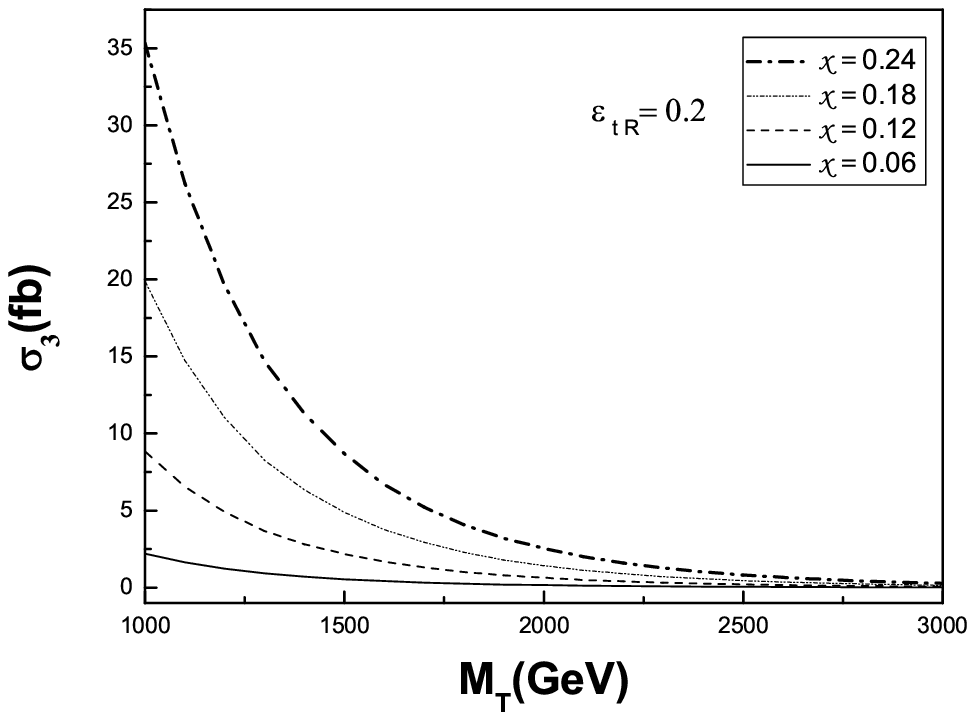,width=200pt,height=180pt} \hspace{-0.5cm}
\hspace{10cm}\vspace{-1cm}
\epsfig{file=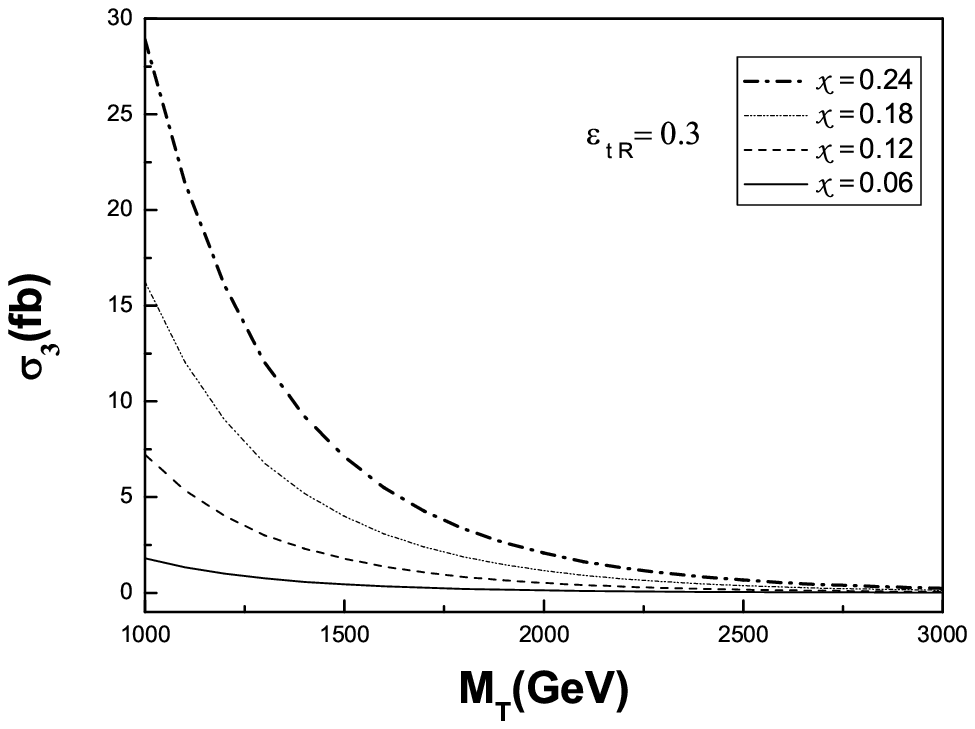,width=200pt,height=180pt} \hspace{-0.5cm}
\vspace{-0.25cm} \put(-110,5){(c)}\put(115,5){ (d)}
\epsfig{file=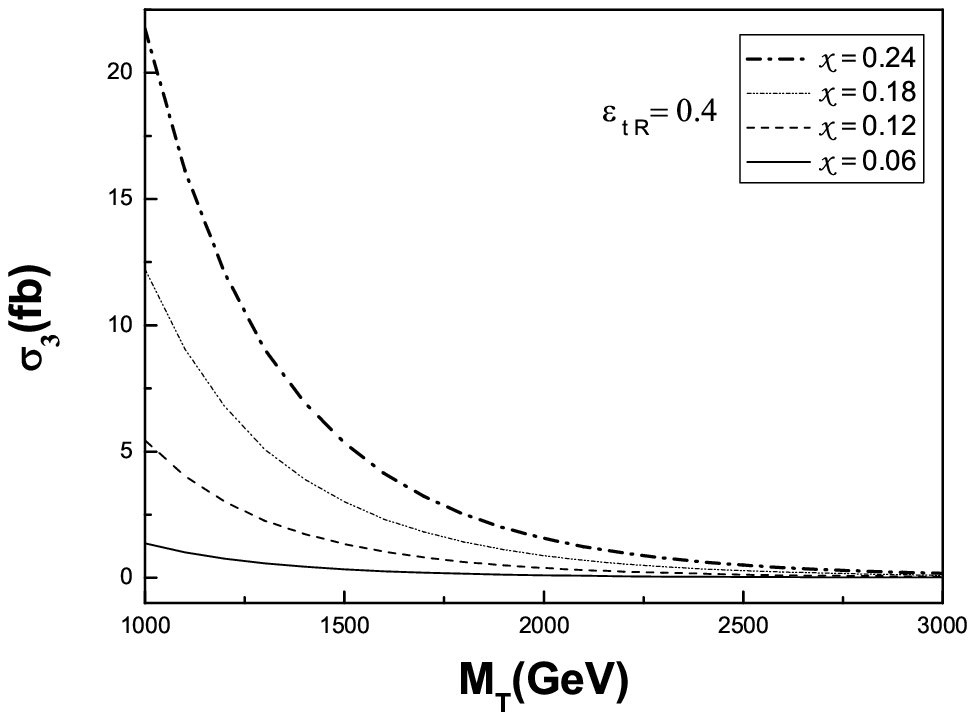,width=200pt,height=180pt} \hspace{-0.5cm}
\hspace{10cm}\vspace{-0.2cm} \caption {Same as Fig.1 but for the
subprocess $qb \rightarrow q'T$ }
\end{center}
\end{figure}

The $q'T$ event can be detected via the
subprocesses:
\begin{equation}
qb\rightarrow q'T\rightarrow q'Zt \rightarrow q'ZWb;
\hspace{2.5cm}qb\rightarrow q'T\rightarrow q'Wb.
\end{equation}
It is well known that the leptonic decay modes of the gauge bosons
$Z$ and $W$ are easier to be identified than the hadronic modes.
Thus, we assume that the gauge bosons $Z$ and $W$ all decay to
leptonic modes i.e. $l^{+}l^{-}$ and $l\bar{\nu}_{l}$. In this case,
the signal events of the above two processes should be chosen as
$q'l^{+}l^{-}l\bar{\nu}_{l}b$ and $q'l\bar{\nu}_{l}b$, respectively.
Both signal events have the spectator jet $q'$ in the forward
direction, which can be utilized to suppress the large $SM$
backgrounds, generated by the $t\bar{t}$, $Wbb$ events [22].

\begin{figure}[htb]
\begin{center}
\epsfig{file=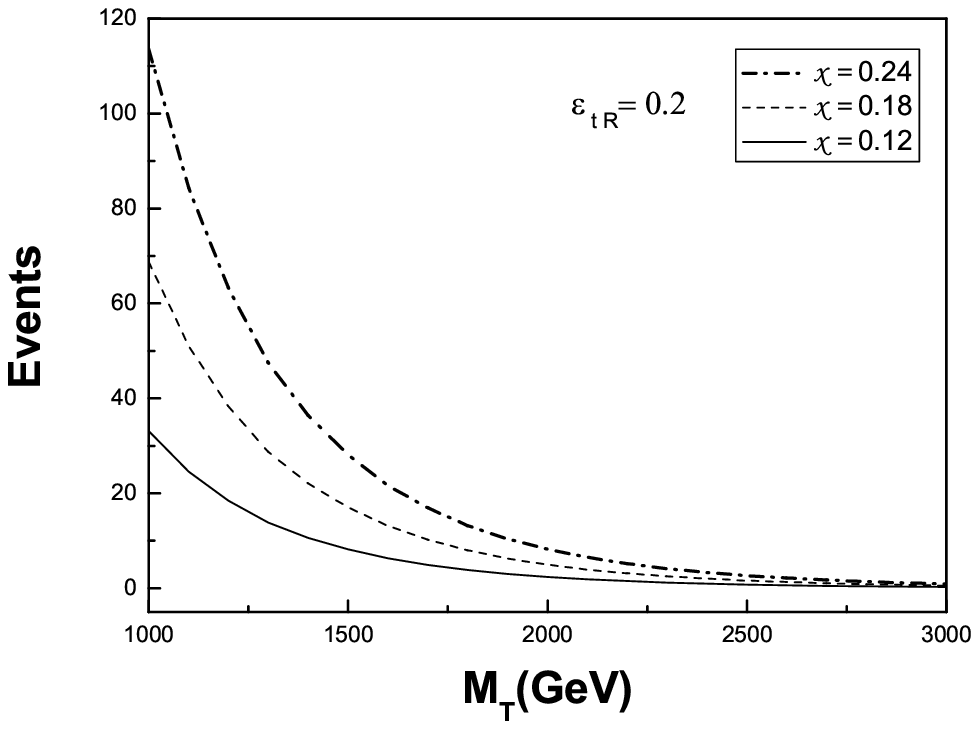,width=220pt,height=205pt} \put(-110,3){
(a)}\put(115,3){ (b)} \vspace{-0.25cm}
\epsfig{file=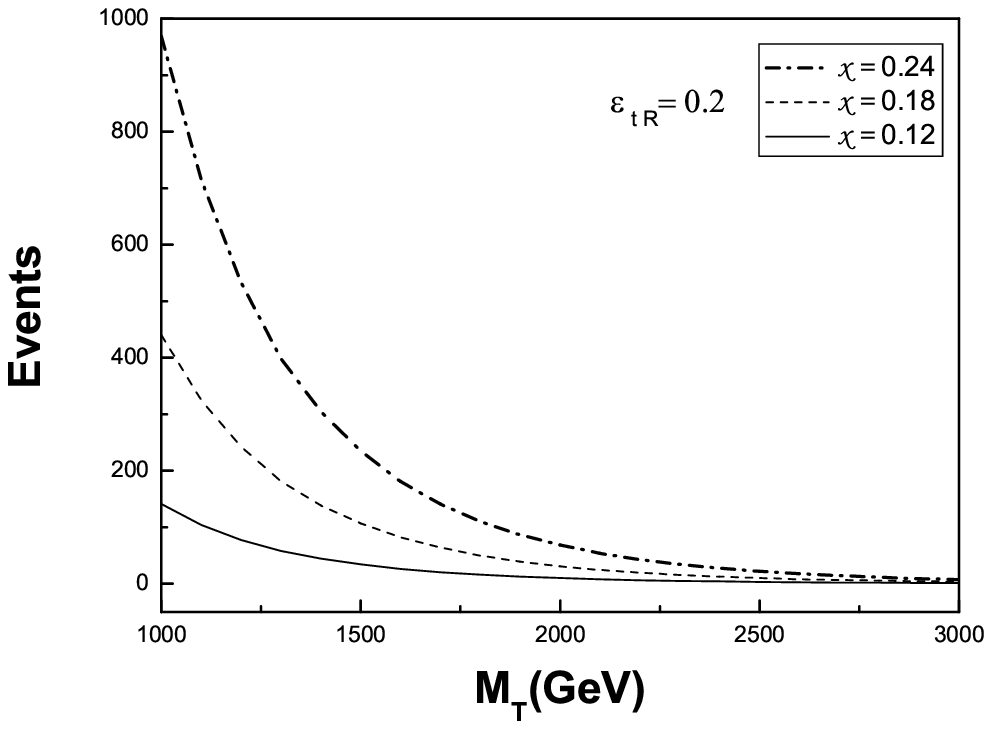,width=220pt,height=207pt} \hspace{-0.5cm}
\hspace{10cm}\vspace{-1cm} \caption{(a) The number of the
$q'l^{+}l^{-}l\bar{\nu}_{l}b$ event as a function of $M_{T}$ for
$\varepsilon_{tR} = 0.2$ and \hspace*{1.9cm}three values of the free
parameter $\chi$ ; (b) Same as $(a)$ but for the $q'l\bar{\nu}_{l}b$
event.} \label{ee}
\end{center}
\end{figure}

The numbers of the signal events $q'l^{+}l^{-}l\bar{\nu}_{l}b$ and
 $q'l\bar{\nu}_{l}b$ at the $LHC$ with $\sqrt{S}=14TeV$
and $\pounds=300fb^{-1}$ are plotted as functions of the $T$ mass
$M_{T}$ in $Fig.5(a)$ and $(b)$, respectively, in which we have
taken $\varepsilon_{tR}=0.2$ and  $\chi$ = 0.12, 0.18, and 0.24.
From these figures, one can see that in most of the parameter space
of the three-site Higgsless model, there will be several tens and up
to thousands of observed $q'l^{+}l^{-}l\bar{\nu}_{l}b$ and
 $q'l\bar{\nu}_{l}b$ events to be generated one year. Thus,
the possible signatures of the heavy top quark $T$ should be
detected via the subprocess $qb\rightarrow q'T$ at the $LHC$
experiments.

Certainly, the heavy top quark $T$ can also be singly produced via
the subprocess $gb\rightarrow TW$, as shown in $Fig.3(c)$ and
$Fig.3(d)$. However, its production cross section is much smaller
than that of the process $qb\rightarrow q'T$. So, it is very
difficult to detect the possible signals of the $T$ quark via the
subprocess $gb\rightarrow TW$.

\noindent{\bf III. Conclusions }

Higgsless model is interesting candidate solution to the puzzle of
$EWSB$. The three-site Higgsless model, which is the simplest
deconstructed Higgsless model, can explain many issues of current
interest in Higgsless models, such as ideal fermion delocalization,
precision electroweak corrections, and fermion mass generation. This
model predicts the existence of the heavy top quark $T$, which might
generate the observed signature in future high energy experiments.

In the context of the three-site Higgsless model, we discuss single
production of the heavy top quark $T$ at the $ILC$, $THERA$, and
$LHC$ experiments. We find that the heavy top quark $T$ can be
produced via the process $e^{+}e^{-}\rightarrow t\overline{T}$ at
the $ILC$ with $\sqrt{S} = 3TeV$. Although, its production cross
section is smaller than $1fb$ in most of the parameter space. The
$SM$ backgrounds can be strongly reduced by applying appropriate
cuts. The possible signals of the heavy top quark $T$ might be
detected via the process $e^{+}e^{-}\rightarrow
t\overline{T}\rightarrow t\bar{t}Z$ in future $ILC$ experiments. The
cross section for the $T$ production associated with an electron
neutrino via the t-channel $W$ exchange process
$eb\rightarrow\nu_{e}T$ at the $THERA$ is larger than that for the
process $e^{+}e^{-}\rightarrow t\overline{T}$ at the $ILC$.
Considering the $THERA$ with $\sqrt{S}=3.7TeV$ and
$\pounds=100pb^{-1}$ has a lower luminosity, there will be a few
$\nu_{e}T$ events to be generated one year. However, enhancing the
value of the integral luminosity of the $THERA$ experiments can
largely increase the number of the $ \nu_{e}T$ events.

The heavy top quark $T$ can be singly produced via the subprocesses
$q\overline{q'}\rightarrow T\bar{b}$, $qb\rightarrow q'T$, and
$gb\rightarrow TW$ at the $LHC$. Our numerical results show that the
production cross sections for the processes
$q\overline{q'}\rightarrow T\bar{b}$ and $gb\rightarrow TW$ are much
small, which can not generate enough number of observed signal
events at the $LHC$ experiments. For the subprocess $qb\rightarrow
q'T$, it is not this case. In most of the parameter space, there
will be several tens of observed $q'l^{+}l^{-}l\bar{\nu}_{l}b$ or
$q'l\bar{\nu}_{l}b$ events to be generated per year. Thus, the
possible signals of the heavy top quark $T$ predicted by the
three-site Higgsless model should be observed via the subprocess
$qb\rightarrow q'T$ at the $LHC$.

The little Higgs models [23] also predict the existence of the heavy
vector-like top quark, its production in future high energy collider
experiments are studied in Refs.[19,24]. They have shown that the
heavy vector-like top quark can also produce the observed signatures
at the $LHC$. Thus, in order to enhance the possibility of detecting
the heavy top quark and to distinguish the three-site Higgsless
model from the little Higgs models, more detailed study is needed.

\vspace{1cm}

\noindent{\bf Acknowledgments}

This work was supported in part by Program for New Century Excellent
Talents in University(NCET-04-0290), the National Natural Science
Foundation of China under the Grants No.10475037 and 10675057.

\end{document}